\documentclass{article}
\usepackage{arxiv}
\usepackage{amsmath}
\usepackage[utf8]{inputenc} 
\usepackage[T1]{fontenc}    
\usepackage{hyperref}       
\usepackage{xcolor}
\definecolor{custom-grey}{RGB}{120,120,120} 
\hypersetup{
    colorlinks=true, 
    linkcolor=blue, 
    filecolor=magenta,      
    urlcolor=custom-grey,
    citecolor=blue}
\usepackage{url}            
\usepackage{booktabs}       
\usepackage{amsfonts}       
\usepackage{nicefrac}       
\usepackage{microtype}      
\usepackage{lipsum}		
\usepackage{graphicx}
\usepackage{natbib}
\usepackage{doi}
\usepackage[margin=10pt,font=tt,labelfont=bf, justification=centerlast]{caption}

\title{Time depending magnetization of nanoparticles under radiofrequency fields: experimental relaxation time in water for solid-liquid transition}

\author{ {\hspace{1mm}Pedro Mendoza Z\'elis$^{1,3}$, Daniel G. Actis$^{1}$, Giuliano A. Basso$^{1,2}$, Gustavo A. Pasquevich$^{1,3}$ and Ignacio J. Bruvera$^{1,2}$ }\thanks{\texttt{pmendoza@fisica.unlp.edu.ar}}\\1: Instituto de Física La Plata (IFLP),UNLP-CONICET.\\ 2: Departamento de Física, Facultad de Ciencias Exactas, UNLP. La Plata, Argentina\\ 3: Departamento de Ciencias Básicas, Facultad de Ingeniería, UNLP, La Plata, Argentina}

\renewcommand{\shorttitle}{\textit{arXiv} Template}

\hypersetup{
pdftitle={A template for the arxiv style},
pdfsubject={q-bio.NC, q-bio.QM},
pdfauthor={David S.~Hippocampus, Elias D.~Striatum},
pdfkeywords={First keyword, Second keyword, More},
}

\begin{document}
\maketitle

\begin{abstract}
In application as hyperthermia and nanowarming, power dissipation arises when the time-dependent magnetization $M(t)$ of an out-of-equilibrium system of nanoparticles lags behind the applied field $H(t)$. The key parameter governing this process is the relaxation time $\tau$ of the system, which induces a phase shift $\phi_n$ between $H(t)$ and every nth harmonic component of $M(t)$.\\
In this work, we present an expression for $M(t)$ in terms of $\tau$ and the equilibrium magnetization, valid for any magnetic system exhibiting odd equilibrium response. From this calculation, we obtain a method for determining the effective $\tau$ of a MNPs sample directly from the experimental measurement of $M(t)$. Additionally, we demonstrate that the power dissipation (SAR: Specific Absorption Rate) of any magnetic sample under a sinusoidal field can be obtained from the first harmonic component of $M(t)$. 
 
As an illustrative application, we explore the variation of $\tau$ for magnetic MNPs in aqueous suspension during the melting process of the matrix. In this case, the change in $\tau$ can be understood as a result of the reorientation of the MNPs in the direction of the applied field as the matrix becomes liquid.

\end{abstract}


\section{Introduction}
The power dissipation of magnetic nanoparticles (MNPs) exposed to radiofrequency fields (RF) enables the development of several applications in biomedicine such as controlled drug release \cite{bruvera2015integrated, salazar2024controlled}, frozen tissue rewarming \cite{han2023vitrification, chen2023nanowarming} and oncological hyperthermia \cite{stephen2021recent}. In all these applications, MNPs absorb energy from the field and release it to their surroundings as heat. The factor of merit for this process is named Specific Absorption Rate (SAR) and expressed as power dissipation per unit mass of magnetic nanoparticles at a given RF field amplitude and frequency. The SAR value of a set of MNPs depends not only on the magnetic properties of the sample and the RF parameters, but also on the supporting medium and the spatial distribution of the particles, which determines the strength of dipolar interactions. \cite{bruvera2019typical}\\
Power dissipation occurs when the magnetization $M(t)$ of a system lags behind the applied field $H(t)$ due to a finite relaxation time $\tau$ which depends on the dispersion medium viscosity and the magnetic properties of the sample. For a sinusoidal magnetic field with amplitude $H_0$ and angular frequency $\omega$, $M(t)$ will also exhibit sinusoidal behavior if the equilibrium magnetic response can be described by $M_{eq}(H)=\chi_0H$, where $\chi_0$ is the equilibrium susceptibility. This linear response occurs for sufficiently low magnetic fields. The sinusoidal response of $M(t)$ will be characterized by an amplitude $M_0$, proportional to $H_0$ and a phase difference $\phi$ with respect to $H(t)$. In this sense, $M(t)$ could be expressed as

\begin{equation}
M(t) =H_0 (\chi' \cos(\omega t) + \chi'' \sin(\omega t))   
\end{equation}

where $\chi’$ is the in-phase component, and $\chi''$ is the out-of-phase component of $\chi$.\\
For large enough $H_0$, the linearity is lost and $M(t)$ will contain odd harmonic components of higher frequency. \cite{drobac2013role, bruvera2022raiders}\\
In a seminal work from Rosensweig \textit{et al.} \cite{rosensweig2002heating} a expression of SAR for a system in the linear regimen is obtained as a function of sample and field parameters, using Shliomis equation for magnetic relaxation

\begin{equation} 
\frac{\partial M(t)}{\partial t}= -\frac{M(t)-M_0(t)}{\tau}\label{schi} 
\end{equation}

, where $M_0(t)$ is the equilibrium magnetization and $\tau$ the effective relaxation time.  
The approximation 
\begin{equation}
	SAR\approx\mu_0 H_0^2 \frac{\omega}{2}\chi_0\frac{\omega \tau}{1+\left(\omega\tau\right)^2}=\mu_0\frac{\omega}{2}  H_0^2 \chi'' \label{Ros}\end{equation}

with $\mu_0$ the vacuum permeability, $\chi_0$ the equilibrium susceptibility per unit mass and $\chi''=\chi_0\frac{\omega \tau}{1+\left(\omega\tau\right)^2}$,  is valid in the linear regimen.\\

Vandendriessche \textit{et al} \cite{vandendriessche2013magneto} demonstrated that, for a system with Langevin equilibrium magnetization, the frequency components of the nonlinear response $M(t)$ for large fields correspond to the odd harmonics of the field frequency. In this work, we generalize this conclusion to any equilibrium magnetic response satisfying $M(H)=-M(-H)$ and to any out-of-equilibrium response to an harmonic magnetic field. Using these results, we are able to determine the effective relaxation time $\tau$ and SAR (specific absorption rate) values of an MNPs sample under an RF sinusoidal field using only the amplitude and phase shift of the first harmonic component of $M(t)$.

As an application example, we obtain the RF magnetic cycles $M(H)$ of an aqueous suspension of MNPs for temperatures in the range of [-20; 20]$^o$C and use them to study the change in $\tau$ during the transition from solid to liquid.

\section{Analytical solution and power dissipation for the out-of-equilibrium magnetization $M(t)$}

\subsection{Fourier expansion for equilibrium magnetization}
For a system magnetically antisymmetric in the direction of an applied field \textit{i.e.} $M(H)=-M(-H)$ with $H$ the field and $M$ the magnetization, the equilibrium magnetization can be written as $M_{eq}(H)=M_s f(H)$ with $M_s$ the saturation value and $f$ a function such satisfying $f(H)=-f(-H)$ and $\lim_{H\to \infty} f(H)= 1$.\\
Expanding $M_{eq}$ in powers of $H$ we get
\begin{equation}
    M_{eq}(H)=M_S\sum\limits_{n=0}^\infty \frac{f^{(2n+1)}(0)}{(2n+1)!}H^{2n+1}
\end{equation}
where only the odd-integer powers of $H$ survive due to the odd nature of $f(H)$.\\

For a harmonic field $H(t)=H_0 \cos (\omega t)$, the equilibrium magnetization at the instant $t$ takes the form 
\begin{equation}
	M_{eq}(H(t))=M_S\sum\limits_{n=0}^\infty \frac{f^{(2n+1)}(0)H_0^{2n+1}}{(2n+1)!}\cos^{2n+1}(\omega t)
\end{equation}

Using Euler’s formula, Newton's binomial theorem and binomial coefficients properties, cosine powers can be rewritten to obtain 
\begin{equation}
	M_{eq}(H(t))=M_S\sum\limits_{n=0}^\infty \frac{f^{(2n+1)}(0)H_0^{2n+1}}{(2n+1)!\,2^{2n}}\sum\limits_{m=0}^n{2n+1\choose n-m} \cos\big((2m+1)\omega t\big)
\end{equation}

that can be rearranged as 
\begin{equation}
	M_{eq}(H(t))=M_S\sum\limits_{p=0}^\infty \alpha_{2p+1} \cos\big((2p+1)\omega t\big)\label{M0}
\end{equation}

what constitutes a common harmonic series with 

\begin{equation}  \alpha_{2p+1}=\sum\limits_{n=p}^\infty \frac{f^{(2n+1)}(0)H_0^{2n+1}}{(2n+1)!\,2^{2n}}{2n+1\choose n-p}\end{equation}

\subsection{Non equilibrium solution}
In order to calculate the non equilibrium magnetization $M(t)$, and taking into account that $M(t)$ has the same period than $H(t)$, we can expand it in a general Fourier series as 
\begin{equation}
	M(t)= \sum\limits_{k=0}^\infty \beta_k \cos\left(k \omega t\right)+\sum\limits_{k=1}^\infty \gamma_k \,\sin(k \omega t)
\end{equation}

, the Shliomis equation for non equilibrium magnetization \ref{schi} can be written as 
\begin{align}
	&\sum\limits_{k=1}^\infty (-1)\beta_k k \omega \,\text{sin}\left(k \omega t\right)+\sum\limits_{k=1}^\infty \gamma_k k \omega \,\cos\left(k \omega t\right)= \nonumber\\
&=-\frac{1}{\tau}\left(\sum\limits_{k=0}^\infty \beta_k \cos\left(k \omega t\right)+\sum\limits_{k=1}^\infty \gamma_k \,\text{sin}\left(k \omega t\right)\right)+\frac{M_S}{\tau}\sum\limits_{p=0}^\infty \alpha_{2p+1} \cos\big((2p+1)\omega t\big)
\end{align}
where equation \ref{M0} has been used, $\tau$ is the effective relaxation time and $M_{eq}(H(t))$ takes the place of $M_0(t)$.\\
Comparing terms in order to get $\beta_k$ and $\gamma_k$ in function of $\alpha_{2p+1}$ and $\tau$, $M(t)$ takes the form
\begin{equation}
	M(t) = M_S\sum\limits_{p=0}^\infty \frac{\alpha_{2p+1}}{1+\left((2p+1) \omega\tau\right)^2}\left\{\cos\big((2p+1)\omega t\big)+ (2p+1) \omega \tau \,\text{sin}\big((2p+1) \omega t\big)\right\}
\end{equation}
which can be simplified using the relation $a \cos(\theta)+b\,\text{sen}(\theta)=R\cos(\theta-\phi)$ with $R=\sqrt{a^2+b^2}$ and $\tan(\phi)=b/a$ to obtain
\begin{equation}
M(t)= M_S\sum\limits_{p=0}^\infty \frac{\alpha_{2p+1}}{\sqrt{1+\left((2p+1) \omega\tau\right)^2}} \cos\big((2p+1) \omega t-\phi_{2p+1}\big)
\label{Mt}
\end{equation}

with $\tan\left(\phi_{2p+1}\right)=(2p+1)\omega\tau$, where it can be seen that the effective relaxation time $\tau$ can be derived from the phase shifts of any odd harmonic term.  In particular, taking the phase $\phi_1$ between the field and the fundamental component of the magnetization ($p=$0), we get 

\begin{equation}
\tau=\frac{\tan(\phi_1)}{\omega}
\label{tau}
\end{equation}

So, Equation \ref{Mt} provides a comprehensive expression for non-equilibrium magnetization in the presence of a sinusoidal field $H(t) = H_0 \cos (\omega t)$, in terms of the odd harmonics of the field frequency $\omega$. It is a general solution, with the sole restriction that the equilibrium magnetization $M_{eq}(H)$ is an odd function. Then, the particular magnetic response of a sample is dictated by the effective $\tau$ and the coefficients $\alpha_{2p+1}$ that derives from $M_{eq}(H)$.

\subsection{SAR calculation}
The Specific Absorption Rate (SAR) can be defined as the magnetic work per unit mass on an adiabatic system, equal to the variation of internal energy per unit mass $\Delta u$, times the field cycling frequency
\begin{equation}
	\text{SAR}=\frac{\omega}{2\pi}\Delta u=-\frac{\omega}{2\pi [MNP]}\mu_0 \oint M dH
 \label{SAR}
\end{equation}

where $M$ is the magnetization per unit mass of the system, $[NPM]$ is the MNP mass concentration and we integrate over a full cycle of the magnetic field $H(t)=H_0\cos(\omega t)$.\\
Making the substitution $\omega t=\theta$ for simplicity,  we can write $dH=-H_0\sin(\theta)d\theta$, and replacing the time dependent magnetization \ref{Mt} into \ref{SAR} we get

\begin{equation}{SAR}=\frac{\omega}{2\pi }\mu_0H_0\frac{M_S}{[NPM]}\sum\limits_{p=0}^\infty\frac{\alpha_{2p+1}}{\sqrt{1+((2p+1)\theta)^2}}\int\limits_0^{2\pi}\cos\left((2p+1)\theta-\phi_{2p+1}\right)\,\text{sen}(\theta)d\theta\end{equation}

Using the sines of the sum and the difference of two angles

\begin{equation}
	\text{SAR}=\frac{\omega}{2\pi }\mu_0 H_0 \frac{M_S}{[NPM]}\sum\limits_{l=1}^\infty c_l  \pi \delta_{p0}\,\text{sen}\left(\phi_l\right)=\mu_0 H_0 M_S \frac{\omega}{2} c_{1}\text{sen}\left(\phi_1\right)
\end{equation}
We can express the sine of the phase as $\sin(\phi_1)=\omega\tau/\sqrt{1+(\omega\tau)^2}$, obtaining for the SAR
\begin{equation}
{SAR}=\mu_0 H_0 \frac{M_S}{[NPM]}\frac{\omega}{2}\frac{\omega\tau}{1+(\omega\tau)^2}\alpha_1
\end{equation}
with 
\begin{equation}
\alpha_1=\sum_{n=0}^{\infty}\frac{f^{(2n+1)}(0)H_0^{2n+1}}{n!(n+1)!2^{2n}}\label{alfa}\end{equation}

remembering the normalized equilibrium magnetization $f(H)=M(H)/M_S$.

This result arise from the fact that the power dissipation of any magnetic system exposed to a sinusoidal magnetic field depends only on the first harmonic component of the magnetization (corresponding to the field frequency) and its phase, and is valid for any $M(t)$. Additionally, the SAR value for any RF field amplitude can be derived from the effective $\tau$ value and $M_{eq}(H)$.\\

\subsection{Magnetization and SAR for a superparamagnetic system}
 For a superparamagnetic system of MNPs with magnetic moment $\mathfrak{m}$ and Langevin response $f(H)=\mathfrak{L}\left(\frac{\mathfrak{m}\mu_0H}{kT}\right)=\mathfrak{L}(x)$ the non zero odd derivatives in \ref{alfa} take the form
\begin{equation}
	f^{(2n+1)}(0)=x^{2n+1}\frac{2^{2n+1}B_{2n+2}}{n+1} 
\end{equation}
so we get for  $\alpha_1$ in \ref{alfa}
\begin{equation}
  \alpha_1^{SPM}=\sum^\infty_{n=0}\frac{2B_{2n+2}}{(2n+2)!}x^{2n+1}{2n+2\choose n+1}
    \label{alfa1}
\end{equation}
where we can express the Bernoulli number in terms of the Riemann Zeta function as
\begin{equation}
   B_{2n+2}=\frac{(-1)^n2(2n+2)!}{(2\pi)^{2n+2}}\zeta(2n+2)
    \label{ber}
\end{equation}
Combining equations \ref{alfa1} and \ref{ber} we obtain
\begin{equation}
\alpha_1^{SPM}=\sum^\infty_{n=0}\frac{(-1)^n\zeta(2n+2)}{2^{2n}\pi^{2n+2}}x^{2n+1}{2n+2\choose n+1}
\end{equation}
where the first terms are
\begin{equation}
   \alpha_1^{SPM}(x)=\frac{x}{3}-\frac{x^3}{60}+\frac{x^5}{756}-\frac{x^7}{8640}+\frac{x^9}{95040}+\mathcal{O}(x^{11})
\end{equation}
where, whit the first term $x/3$ we retrieve equation \ref{Ros} for linear magnetization.\\

For a more convenient expression where the degree of approximation does not depends on the power of $x$, we can write the Zeta function 
\begin{equation}
	\zeta(s)=\sum\limits_{k=1}^\infty\frac{1}{k^s}
\end{equation}
and exchanging the sums 
\begin{equation}
    \alpha_{1}^{FF}=4 \sum\limits_{k=1}^\infty \sum\limits_{n=0}^\infty \frac{(-1)^{n}}{\left(2 k\pi\right)^{2n+2}} x^{2n+1}{2n+2\choose n+1}=\frac{4}{x} \sum\limits_{k=1}^\infty\left(1-\frac{1}{\sqrt{1+\frac{x^2}{\pi^2 k^2}}}\right)
    \label{suma}
\end{equation}

which leads to an asymptotic value of $4/\pi\approx1.273$.\\

\section{Experimental determination of the effective relaxation time evolution during matrix melting}
As a demonstration of $\tau$ determination, we show results for the solid to liquid transition of a MNPs aqueous suspension.

\subsection{Materials and methods}
\subsubsection{Magnetic nanoparticles synthesis and suspension}
Fe$_3$O$_4$ nanoparticles 10(3) nm diameter (TEM) were synthesized by a co-precipitation method of Fe$^{2+}$ and Fe$^{3+}$ salts into magnetite phase followed by citric acid adsorption for achieving electrostatic stability in suspension, preventing aggregation and oxidation, as previously described in \cite{de2013stability}. The particles were suspended in distilled water at a concentration of 7.4(5) g/L determined using UV-Vis spectroscopy, employing a method derived from \cite{adams1995determining}.

\subsubsection{RF cycles measurement}
The RF field was generated by a power source-resonator set \textit{Hüttinger TIG 2,5/300} with a [30; 300] kHz nominal frequency range and a 2.5 kW maximum output, reaching a maximum field amplitude of 54 kA/m with a water refrigerated 6 turn coil of 2.5 cm diameter.\\

RF magnetization cycles were obtained for all samples with an inductive  system as reported in \cite{bruvera2022raiders}. The system was calibrated using a $Dy_2O_3$ paramagnetic pattern. The temperature of the samples was measured by a Rugged LSENS-T optical probe inserted in the center of the fluid. \\
The signal processing used for obtainig the magnetic cycles relies on the application of a Fourier Transform. Voltage signals corresponding to $\partial H/\partial t$ and $\partial M/\partial t$ are acquired, numerically integrated and then subjected to the FFT algorithm, resulting in frequency and phase spectra. The field sinusoidal signal exclusively comprises the fundamental harmonic component. Subsequently, the frequency spectrum is filtered to retain only the odd multiples of the fundamental frequency, and $M(t)$ is reconstructed using the inverse FFT algorithm. RF magnetic cycles were then constructed from the $H(t)$ and $M(t)$ signals, and the phase shift was determined from the FFT phase values of the fundamental frequency.\\

The samples were cooled below -20°C using a cooling spray. They were then exposed to the RF field while monitoring their temperature within the ESAR system. Approximately one cycle per second was acquired during the heating process as an average of several measurements. Three experiments were performed for each sample under every field condition. The reported values represent the average of these three measurements and the error bars represent one standard deviation.

\section{Results}
Figure 1 shows the obtained magnetic cycles of the MNPs suspension for temperatures -20(1)ºC and 20(1)ºC together with the effective relaxation time $\tau$ values for the whole temperature range obtained using equation \ref{tau}. A transition around the melting point of water is observed, in which $\tau$ changes approximately from 0.9 x $10^{-7}$ s for $T<-5^{\circ}C$ to 1.4 x $10^{-7}$ s for $T>5^{\circ}C$.
\begin{figure}[ht]
\centering
\includegraphics[width=1\linewidth]{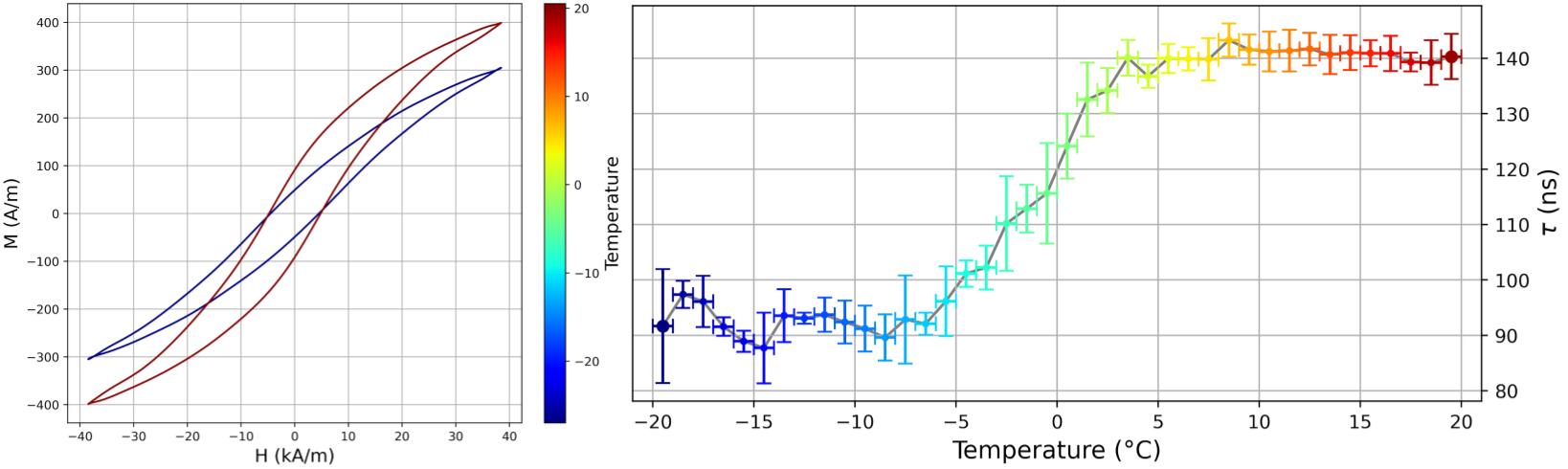}
\label{graf}
\caption{Left: Typical RF magnetic cycles for a 38(1) kA/m, 135(1) kHz field at temperatures T=-20(1)ºC and T=20(1)ºC. Right: Effective relaxation time $\tau$ vs. temperature for T$\in$[-20(1), 20(1)]ºC.}
\end{figure}

The measured cycles are smooth and repetitive with good signal-to-noise ratio. The obtained $\tau$ values are compatibles with the anisotropy constant reported for these MNPs in \cite{de2013stability}.\\ 

\section{Discussion and conclusions}

We studied the evolution of $\tau$ for a sample of MNPs suspended in water during the solid-liquid transition of the matrix. Several experimental runs were conducted, all yielding consistent results. The obtained cycles are clean and exhibit a monotonic evolution of all their parameters between the solid and liquid states. A detailed study of this process will be the focus of future work. The evolution of $\tau$, in particular, shows a clear transition around the melting point of water, from a value of approximately 90 ns for temperatures below -5$^{\circ}$ C to a constant value of approximately 140 ns for temperatures above 5$^{\circ}$ C, much larger than predicted from numerical simulations of similar particles in \cite{davidson2024field}. This change can be interpreted as a consequence of the alignment of the anisotropy axes of the MNPs with the field $H(t)$ as proposed in \cite{shah2015mixed}: since the sample was frozen without an applied field, the anisotropy axes are randomly oriented in the solid state. When the aqueous matrix becomes liquid, the particles are free to rotate and align their anisotropy axes with the external field, minimizing energy. The size of the MNPs and previously published results indicate that the dominant relaxation mechanism in liquid water is Nèel relaxation \cite{de2013stability, lima2014relaxation}, where the magnetic moment flips within the particles following the external field, which in this case oscillates between the two opposite directions of the generating coil axis. The energy barrier for the moment to flip between these two directions is further conditioned by the direction of the particles' easy magnetization axis with respect to the applied field. When this axis is perpendicular to the field, there is no barrier, while the maximum value is reached when the easy axis is parallel to the field. Thus, the population of MNPs, initially exhibiting a random distribution of easy axis orientations, ends up with all particles aligned in the direction of the field as the matrix becomes liquid, maximizing the effective relaxation time. These results support the simulations previously published in \cite{chalifour2021magnetic}. 

As the main result of this work, we have derived an expression for the out-of-equilibrium magnetization as a function of time $M(t)$, valid for any magnetic system exhibiting an odd equilibrium response $M_{eq}(H) = -M_{eq}(-H)$. This was achieved starting from the Schliomis equation and expressing $M(t)$ as a Fourier series expansion. With the obtained expression, it is possible to construct $M(t)$ for any amplitude $H_0$ and frequency $\omega$ of the applied field, based on $M_{eq}(H)$ and the effective relaxation time $\tau$. Consequently, by experimentally determining $M(t)$ for a given $H(t)$, it is possible to find the value of $\tau$ as a function of the phase difference between $H(t)$ and any harmonic component of $M(t)$, particularly the fundamental, which has the highest amplitude. With these results, we further demonstrate that the power dissipated by the system under a sinusoidal field depends solely on the amplitude and phase shift of the first harmonic of $M(t)$.\\
The obtained expressions allow for a detailed study of the effective relaxation time evolution of nanoparticles based on RF magnetic cycle records. With this information, the power dissipated by the system can be obtained from the amplitude of the fundamental component of $M(t)$. Understanding the evolution of these quantities in the transition from cryogenic temperatures to room temperature constitutes a powerful tool in the development of the application of MNPs in tissue thawing and oncological hyperthermia.

\bibliographystyle{unsrtnat}
\bibliography{references}  

\end{document}